\documentclass[aps,prx,twocolumn, superscriptaddress,amsmath,amssymb,longbibliography]{revtex4-2}

\usepackage{xcolor}
\usepackage{amsmath}
\usepackage{graphicx}
\usepackage{dcolumn}
\definecolor{webblue}{HTML}{1111aa}
\usepackage[
colorlinks=true,
linkcolor=webblue,
citecolor=webblue,
urlcolor=webblue,
]{hyperref}
\usepackage{bm}
\usepackage[T1]{fontenc}

\begin{document}

\preprint{APS/123-QED}

\title{Distinct lattice and charge excitations in \textit{A}V$_{3}$Sb$_{5}$ kagome superconductors}

\author{Dongjin Oh}
\email[Corresponding Author:~]{djeeoh@gmail.com}
\affiliation{Department of Physics, Massachusetts Institute of Technology, Cambridge, MA 02139, USA}

\author{Stefan Enzner}
\affiliation{Institut für Theoretische Physik und Astrophysik and Würzburg-Dresden Cluster of Excellence ct.qmat, Universität Würzburg, 97074 Würzburg, Germany}

\author{Lennart Klebl}
\affiliation{Institut für Theoretische Physik und Astrophysik and Würzburg-Dresden Cluster of Excellence ct.qmat, Universität Würzburg, 97074 Würzburg, Germany}
\affiliation{Institute for Theoretical Physics, Universität Hamburg, Notkestraße 9–11, 22607 Hamburg, Germany}

\author{Harley D. Scammell}
\affiliation{School of Mathematical and Physical Sciences, University of Technology Sydney, Ultimo, NSW 2007, Australia}

\author{Julian Ingham}
\affiliation{Department of Physics, Columbia University, New York, NY, 10027, USA}

\author{Tim Wehling}
\affiliation{Institute for Theoretical Physics, Universität Hamburg, Notkestraße 9–11, 22607 Hamburg, Germany}
\affiliation{The Hamburg Centre for Ultrafast Imaging, Luruper Chaussee 149, 22761 Hamburg, Germany}

\author{Giorgio Sangiovanni}
\affiliation{Institut für Theoretische Physik und Astrophysik and Würzburg-Dresden Cluster of Excellence ct.qmat, Universität Würzburg, 97074 Würzburg, Germany}

\author{Ronny Thomale}
\affiliation{Institut für Theoretische Physik und Astrophysik and Würzburg-Dresden Cluster of Excellence ct.qmat, Universität Würzburg, 97074 Würzburg, Germany}

\author{Ahmet Kemal Demir}
\affiliation{Department of Physics, Massachusetts Institute of Technology, Cambridge, MA 02139, USA}

\author{Connor Occhialini}
\affiliation{Department of Physics, Massachusetts Institute of Technology, Cambridge, MA 02139, USA}

\author{Dirk Wulferding}
\affiliation{Department of Physics and Astronomy, Sejong University, Seoul 05006, Republic of Korea}

\author{Ziqiang Wang}
\affiliation{Department of Physics, Boston College, Chestnut Hill, MA, USA}

\author{Andrea C. Salinas}
\affiliation{Materials Department, Materials Research Laboratory, and California Nano Systems Institute, University of California Santa Barbara, Santa Barbara, California 93106, USA}

\author{Stephen D. Wilson}
\affiliation{Materials Department, Materials Research Laboratory, and California Nano Systems Institute, University of California Santa Barbara, Santa Barbara, California 93106, USA}

\author{Riccardo Comin}
\email[Corresponding Author:$~$]{rcomin@mit.edu}
\affiliation{Department of Physics, Massachusetts Institute of Technology, Cambridge, MA 02139, USA}

\begin{abstract}
The kagome superconductor family $\textit{A}$V$_{3}$Sb$_{5}$ ($\textit{A}$ = Cs, Rb, K) provides a rich platform for exploring diverse electronic symmetry breaking phenomena, including superconductivity and various forms of density wave orders. Although these compounds share the identical lattice structure in the normal state, they exhibit distinct forms of symmetry breaking upon entering the charge density wave (CDW) phase, and the microscopic origin of which remain elusive. Here, we investigate the lattice and charge degrees of freedom in \textit{A}V$_{3}$Sb$_{5}$ using angle-resolved polarized Raman spectroscopy. Our comprehensive polarization-resolved measurements reveal that the lifting of the twofold-degeneracy of the $E_{2g}$ phonon mode in the CDW phase---previously reported only in CsV$_{3}$Sb$_{5}$ with a 3 GHz splitting---also appears ubiquitously in the other two compounds. In contrast, the collective CDW excitations exhibit markedly different polarization dependences depending on the alkali-metal species. These distinct behaviors in the lattice and charge channels provide crucial insight into the enigmatic material-dependent symmetry breaking phenomena that appear in the CDW phase. Furthermore, our experiments, together with first-principles calculations and an effective Hamiltonian model, shed light on the nature of the charge order structure in $\textit{A}$V$_{3}$Sb$_{5}$ kagome superconductors.
\end{abstract}

\maketitle

\textit{Introduction}---The kagome superconductors CsV$_{3}$Sb$_{5}$, RbV$_{3}$Sb$_{5}$, and KV$_{3}$Sb$_{5}$ have gained significant attention as promising material platforms for exploring exotic broken-symmetry phases~\cite{wilson_av3sb5_2024}. All three compounds exhibit both superconductivity and CDW phases, although the exact nature of the charge order structures remains elusive~\cite{ortiz_new_2019,ortiz_csv3sb5_2020,kautzsch_structural_2023}. Furthermore, time-reversal and rotational symmetry breakings have been reported below the CDW transition temperature, \textit{T}$_{\text{CDW}}$, in all three materials~\cite{khasanov_time-reversal_2022,mielke_time-reversal_2022,guguchia_tunable_2023,li_rotation_2022,li_unidirectional_2023,xu_three-state_2022}. 

Despite sharing identical crystalline structures in the normal phase, the $\textit{A}$V$_{3}$Sb$_{5}$ compounds exhibit distinct symmetry breaking within the CDW phase. Notably, a lattice anomaly has been reported exclusively in CsV$_{3}$Sb$_{5}$~\cite{wu_charge_2022,wulferding_emergent_2022,ratcliff_coherent_2021}, while scanning tunneling microscopy experiments have further revealed additional rotational symmetry breaking associated with a 4\textit{a}$_{0}$ unidirectional charge stripe order in CsV$_{3}$Sb$_{5}$ and RbV$_{3}$Sb$_{5}$, a feature notably absent in the KV$_{3}$Sb$_{5}$~\cite{zhao_cascade_2021,li_small_2023,xing_optical_2024}. Additionally, CsV$_{3}$Sb$_{5}$ displays electronic magnetochiral anisotropy supporting the presence of chiral electronic order, which is also absent in KV$_{3}$Sb$_{5}$~\cite{guo_switchable_2022, guo_distinct_2024}.

\begin{figure}
	\includegraphics[width=5.5cm]{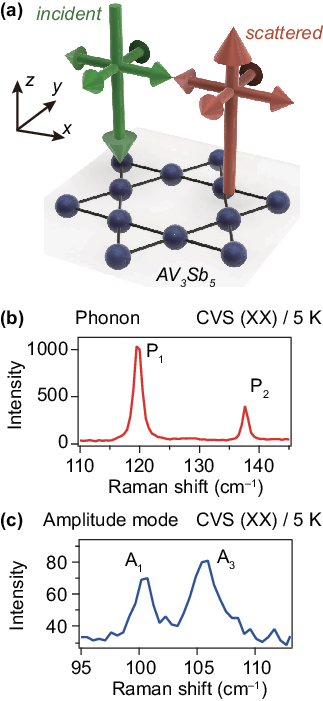}
	\caption{    
		(a) Experimental geometry of polarization-resolved Raman spectroscopy. Green and red arrows indicate incident and scattered lights, respectively.
            (b) Phonon spectrum of CsV$_{3}$Sb$_{5}$ measured in parallel polarization (XX) geometry at 5 K.
            (c) CDW amplitude mode spectrum of CsV$_{3}$Sb$_{5}$ measured in XX geometry at 5 K. 
	}
    \label{fig:1}
\end{figure}

In spite of these intriguing observations, the mechanisms underlying these distinct electronic symmetry breakings and charge order ground state remain poorly understood amidst a number of theoretical proposals~\cite{feng_chiral_2021,tan_charge_2021,scammell_chiral_2023-1,ingham2025vestigial,wang_structure_2023}. Therefore, to fully capture the complex and diverse symmetry-breaking phenomena in the CDW phase of $\textit{A}$V$_{3}$Sb$_{5}$ kagome superconductors, it is essential to systematically analyze the differences arising from their fundamental degrees of freedom.

In this letter, we employ polarization-resolved Raman spectroscopy to selectively investigate the lattice and charge degrees of freedom in the \textit{A}V$_{3}$Sb$_{5}$ kagome superconductors. By tracking collective lattice excitations (phonons), we find the ubiquitous \textit{E}$_{2g}$ phonon splitting in the CDW phase of Cs, Rb, and K compounds. In contrast, the CDW amplitude modes---associated with excitations of the CDW order parameter (modulated charge density)---exhibit distinct angular Raman intensity patterns in KV$_{3}$Sb$_{5}$ that differ markedly from those in CsV$_{3}$Sb$_{5}$ and RbV$_{3}$Sb$_{5}$. These distinct behaviors observed in the collective excitations of lattice and charge degrees of freedom offer insight into possible scenarios for understanding the alkali metal-dependent symmetry breaking phenomena. Moreover, our experiments, in conjunction with first-principles calculations and effective Hamiltonian modeling, illuminate the nature of the charge order structure in \textit{A}V$_{3}$Sb$_{5}$ kagome superconductors.


To independently investigate the lattice and charge degrees of freedom in \textit{A}V$_{3}$Sb$_{5}$, we performed polarized-Raman spectroscopy [Fig.~\ref{fig:1}(a)]. The Raman spectra of \textit{A}V$_{3}$Sb$_{5}$ reveal both the phonon modes and CDW amplitude modes. As shown in Fig.~\ref{fig:1}(b), the Raman spectrum of CsV$_{3}$Sb$_{5}$ measured in parallel (XX) polarization geometry displays two prominent phonon peaks. In the normal state, P$_{1}$ and P$_{2}$ phonons correspond to the $E_{2g}$ doublet and $A_{1g}$ singlet of the $D_{6h}$ point group, respectively \cite{liu_observation_2022,wu_charge_2022,wulferding_emergent_2022}. In addition, XX Raman spectrum of CsV$_{3}$Sb$_{5}$ reveals multiple CDW amplitude modes in the CDW phase as shown in Fig.~\ref{fig:1}(c)~\cite{liu_observation_2022,wu_charge_2022,wulferding_emergent_2022,jin__2024}. To obtain deeper insights into the lattice and charge excitations, we further carried out systematic angle-resolved polarized Raman spectroscopy (ARPRS) experiments.

\begin{figure}[htbp!]
	\includegraphics[width=8.5cm]{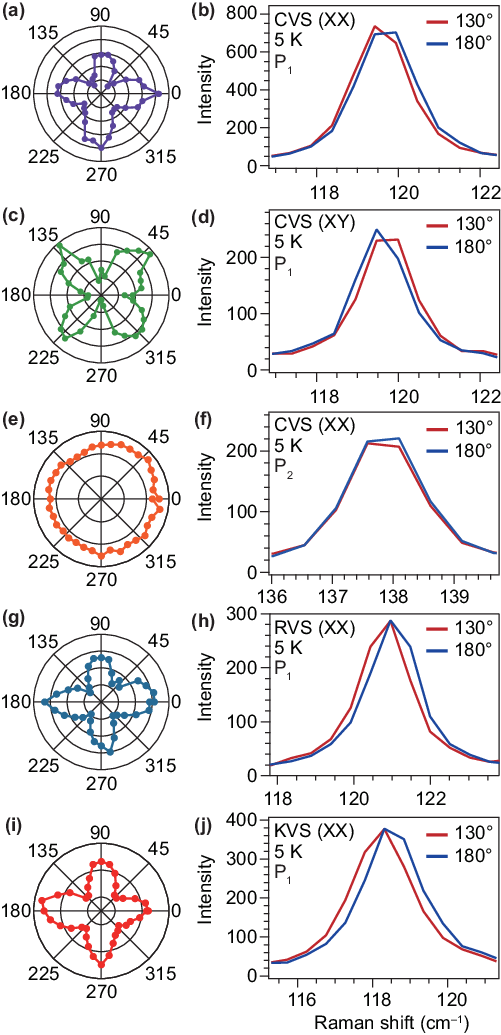}
	\caption{
        Polarization angle ($\theta$)-dependent phonon frequencies of \textit{A}V$_{3}$Sb$_{5}$.
        (a) $\theta$-dependence of the P$_{1}$ phonon frequency in CsV$_{3}$Sb$_{5}$. 
        (b) Raw P$_{1}$ phonon spectra measured at $\theta$ = 130$^{\circ}$ (red) and $\theta$ = 180$^{\circ}$ (blue). 
        (c,d) Corresponding $\theta$-dependent P$_{1}$ phonon frequency and raw spectra from the XY spectrum of CsV$_{3}$Sb$_{5}$. 
        (e,f) $\theta$-dependence of P$_{2}$ phonon frequency in CsV$_{3}$Sb$_{5}$, along with raw P$_{2}$ phonon spectra measured in XX geometry.
        (g-j) $\theta$-dependence of the P$_{1}$ phonon frequency and raw phonon spectra for RbV$_{3}$Sb$_{5}$ (g,h) and KV$_{3}$Sb$_{5}$ (i,j), respectively. All data were collected using a 2.33 eV photon energy.
        }
        \label{fig:2}
\end{figure}

\begin{figure*}[htbp!]
	\includegraphics[width=16.5cm]{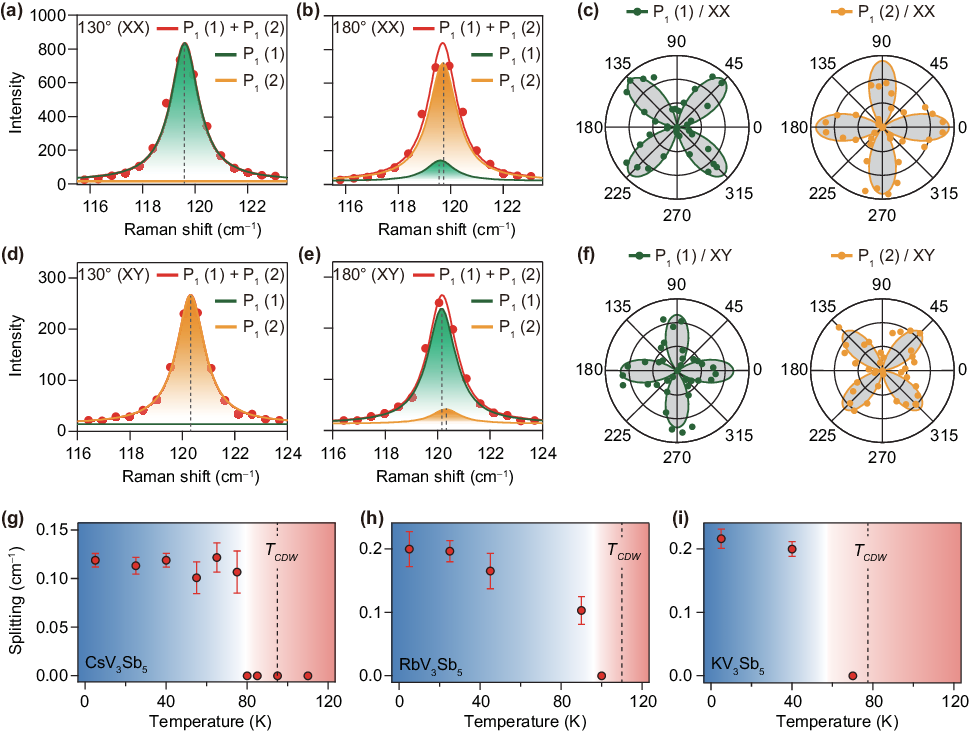}
	\caption{P$_{1}$ phonon splitting of $\textit{A}$V$_{3}$Sb$_{5}$ in CDW phase
		(a,b) Two Lorentzian function fitting results of XX Raman spectra of CsV$_{3}$Sb$_{5}$ measured at $\theta$ = 130$^\circ$ (a) and $\theta$ = 180$^\circ$ (b). The green and orange shaded areas indicate the spectral weights of the P$_{1}$(1) and P$_{1}$(2) phonons, respectively. The red solid curves and circles correspond to the integrated spectral weights of P$_{1}$(1) and P$_{1}$(2) phonons and the raw Raman data, respectively.	
		(c) $\theta$-dependent spectral weights of the P$_{1}$(1) and P$_{1}$(2) phonons, extracted from XX Raman spectra. Circles and solid curves represent experimental data and simulated results, respectively.     
		(d-f) Corresponding plots for the XY Raman spectra of CsV$_{3}$Sb$_{5}$. All data were collected using a 2.33 eV photon energy.
		(g-i) Temperature-dependent P$_{1}$ phonon splitting in \textit{A}V$_{3}$Sb$_{5}$. Black dashed lines mark the $T_{CDW}$ for each material. Error bars reflect the standard deviation of the fitting results.
	}
	\label{fig:3}
\end{figure*}

\textit{Polarization angle-dependent P$_{1}$ phonon frequency modulation}---Figure~\ref{fig:2} illustrates the low-temperature polarization-dependence of the P$_{1}$ and P$_{2}$ phonon frequencies in CsV$_{3}$Sb$_{5}$, RbV$_{3}$Sb$_{5}$, and KV$_{3}$Sb$_{5}$. To capture this dependence, we measured the Raman spectra at various polarization angles ($\theta$) by rotating the polarization of the light while maintaining either XX or cross (XY) polarization geometry. The phonon frequencies were then extracted by fitting the phonon Raman spectra using a single Lorentzian function. The $\theta$-dependence of the P$_{1}$ phonon frequency in CsV$_{3}$Sb$_{5}$ measured in XX geometry shows a clear fourfold modulation consistent with previous study [Fig.~\ref{fig:2}(a)]~\cite{wulferding_emergent_2022}. This frequency modulation is further corroborated by the comparison of raw P$_{1}$ phonon spectra measured at $\theta$ = 130$^\circ$ (red) and $\theta$ = 180$^\circ$ (blue) as shown in Fig.~\ref{fig:2}(b) (see also Supplemental Material). We further extended our investigation by performing the same measurements on CsV$_{3}$Sb$_{5}$ in XY geometry. The XY Raman spectra also exhibit a fourfold frequency modulation as a function of $\theta$. Notably, this modulation is phase-shifted by 45$^\circ$ relative to the XX spectra [Figs.~\ref{fig:2}(c) and~\ref{fig:2}(d)]. In contrast to the P$_{1}$ phonon, no discernible frequency modulation was observed in the P$_{2}$ phonon as shown in Figs.~\ref{fig:2}(e) and~\ref{fig:2}(f). Remarkably, similar fourfold phonon frequency modulations were also observed in RbV$_{3}$Sb$_{5}$ and KV$_{3}$Sb$_{5}$. The fourfold frequency modulation and corresponding spectral shifts for RbV$_{3}$Sb$_{5}$ and KV$_{3}$Sb$_{5}$ measured in XX polarization geometry are presented in Figs.~\ref{fig:2}(g) and~\ref{fig:2}(h), and Figs.~\ref{fig:2}(i) and~\ref{fig:2}(j), respectively. This finding clearly demonstrates that the anomaly observed in the $E_{2g}$ phonon of CsV$_{3}$Sb$_{5}$ is also ubiquitously present in RbV$_{3}$Sb$_{5}$ and KV$_{3}$Sb$_{5}$.

It is important to note that the $\theta$ of the incident and scattered light cannot affect the phonon mode frequency. Furthermore, we can exclude any $\theta$-dependent spectrometer misalignment as a source of the frequency shifts, as the latter would otherwise also manifest in the P$_{2}$ phonon, which, however, has a fixed frequency. Therefore, the observed frequency modulations as a function of $\theta$ can only be explained by a spectral overlap between two independent phonon modes with sub-linewidth splitting and distinct dependence on $\theta$.

\textit{P$_{1}$ phonon splitting in CDW phase}---To disentangle the contributions of two distinct phonon spectral weights, we fit the angle-resolved polarized Raman spectra of CsV$_{3}$Sb$_{5}$ using two Lorentzian functions. During the fitting process, the peak positions of each Lorentzian was fixed to the local minimum and maximum frequencies determined in the previous analysis (Fig.~\ref{fig:2}). The spectral fitting results for the XX Raman spectra of CsV$_{3}$Sb$_{5}$ are illustrated in Figs.~\ref{fig:3}(a) and~\ref{fig:3}(b). At $\theta$ = 130$^\circ$, the raw P$_{1}$ phonon spectrum was well fitted by the lower-energy Lorentzian component, P$_{1}$(1) (green curve and shaded area), with no noticeable contribution from the higher-energy component, P$_{1}$(2) (orange curve). Conversely, at $\theta$ = 180$^\circ$, P$_{1}$(2) dominates the spectral weight, while P$_{1}$(1) contributes only marginally [Fig.~\ref{fig:3}(b)]. By fitting the XX Raman spectra measured across $\theta$ = 0$^\circ$ $\sim$ 360$^\circ$, we extracted the spectral weights of P$_{1}$(1) and P$_{1}$(2). As illustrated in Fig.~\ref{fig:3}(c), the angular intensity patterns of P$_{1}$(1) and P$_{1}$(2) display clear fourfold patterns with a 45$^\circ$ phase shift. In the XY geometry, the behavior of P$_{1}$(1) and P$_{1}$(2) phonon components is reversed. The spectrum at 130$^\circ$ is dominated by P$_{1}$(2), while at 180$^\circ$, P$_{1}$(1) becomes the primary contributor [see Figs.~\ref{fig:3}(d) and~\ref{fig:3}(e)]. This reversed $\theta$-dependence is visualized in the polar plots in Fig.~\ref{fig:3}(f), clearly highlighting the contributions of P$_{1}$(1) and P$_{1}$(2). This distinct behavior unambiguously demonstrates that the split P$_{1}$(1) and P$_{1}$(2) phonons correspond to the $B_{1g}$ and $A_{g}$ modes, respectively, within the $D_{2h}$ notation, in agreement with previous Raman study \cite{jin__2024}. Interestingly, we find that the splitting of the P$_{1}$ phonon ($\Delta$ = 0.1 $\sim$ 0.2 cm$^{-1}$ $\approx$ 3 GHz) is exclusively observed within the CDW phase of \textit{A}V$_{3}$Sb$_{5}$ [Figs.~\ref{fig:3}(g)-(i)]. This temperature-dependent behavior of the P$_{1}$ phonon anomaly indicates the lifting of the $E_{2g}$ phonon degeneracy due to symmetry breaking in the lattice degree of freedom associated with CDW state. From group-theoretical considerations, only breaking of the $C_{3}$ rotational symmetry can lift the degeneracy of the $E_{2g}$ mode in the $D_{6h}$ point group (see Supplemental Material). Therefore, the observed $\theta$-dependent Raman intensity patterns provide unequivocal evidence for the breakdown of $\textit{E}_{2g}$ degeneracy due to rotational symmetry breaking across the CDW transition. 

\begin{figure*}
	\includegraphics[width=16.5cm]{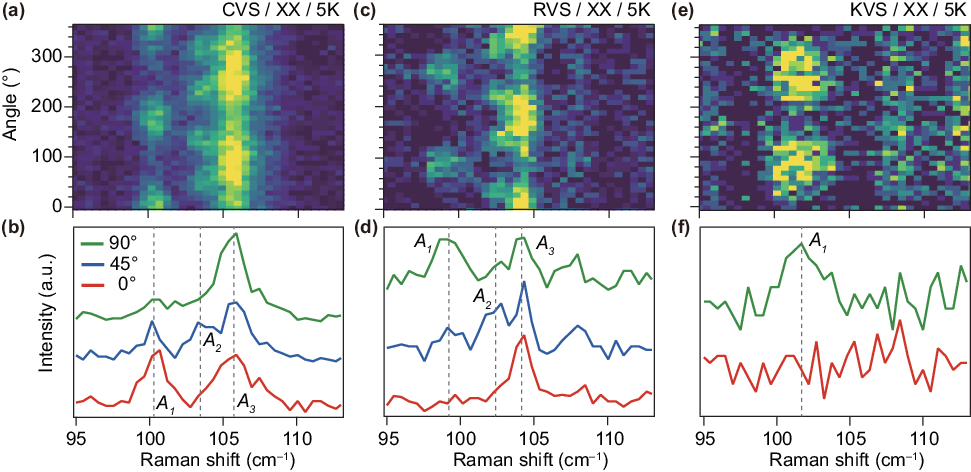}
	\caption{
		Amplitude mode spectra of \textit{A}V$_{3}$Sb$_{5}$.
		(a) Angle-resolved polarized Raman spectra of CsV$_{3}$Sb$_{5}$. 
		(b) Raw Raman spectra of CsV$_{3}$Sb$_{5}$. Red, blue and green curves correspond to measurements at polarization angles ($\theta$) of 0$^\circ$, 45$^\circ$, and 90$^\circ$, respectively. Dashed lines indicate the peak positions of the amplitude modes.
		(c-f) Corresponding plots for RbV$_{3}$Sb$_{5}$ (c and d) and KV$_{3}$Sb$_{5}$ (e and f), respectively. The $\theta$-dependent Raman spectra were measured in the XX polarization geometry with 2.33 eV photon energy.
	}
	\label{fig:4}
\end{figure*}

It is worth noting that our domain-resolved measurements revealed the existence of three types of domains, each exhibiting a fourfold frequency modulation of the P$_{1}$ phonon frequency with a relative phase shift of 120$^\circ$, naturally implying a rotational symmetry breaking from $C_{6}$ to $C_{2}$ in lattice degree of freedom (see Supplemental Material). This observation is reminiscent of previous scanning optical measurements, further supporting the symmetry breaking induced P$_{1}$ phonon splitting in the CDW state~\cite{xu_three-state_2022}. Taken together, our systematic polarization- and domain-resolved Raman measurements provide clear evidence of lattice anomaly observed in all three \textit{A}V$_{3}$Sb$_{5}$ compounds, and suggest that the $E_{2g}$ phonon splitting is closely tied to the ubiquitous rotational symmetry breaking in the lattice sector associated with the CDW state in \textit{A}V$_{3}$Sb$_{5}$.

\textit{Polarization angle-dependent amplitude mode spectra in \textit{A}V$_{3}$Sb$_{5}$}---We further investigated the $\theta$-dependent amplitude mode spectra of \textit{A}V$_{3}$Sb$_{5}$. Fig.~\ref{fig:4}(a) shows the $\theta$-dependent amplitude mode spectra of CsV$_{3}$Sb$_{5}$ measured in the XX geometry from $\theta$ = 0$^\circ$ to $\theta$ = 360$^\circ$. Within the wavenumber range of 95 cm$^{-1}$ to 110 cm$^{-1}$, three amplitude modes, A$_{1}$ (100.5 cm$^{-1}$), A$_{2}$ (103.3 cm$^{-1}$), and A$_{3}$ (105.6 cm$^{-1}$) were identified as shown in Fig.~\ref{fig:4}(b). These CDW amplitude modes reveal distinct $\theta$-dependent intensities. As illustrated in Fig.~\ref{fig:4}(a), the A$_{1}$ and A$_{3}$ amplitude modes exhibit a twofold angular dependence, while the A$_{2}$ amplitude mode displays fourfold angular intensity pattern, consistent with a prior study (Supplemental Material) \cite{jin__2024}. We note that the twofold $\theta$-dependence observed in the amplitude-mode spectra already reflects the underlying $C_{2}$ rotational symmetry. This indicates that the charge degree of freedom in the CDW phase is likewise governed by the rotational-symmetry breaking in line with the behavior of lattice sector. Interestingly, consistent amplitude mode spectra and polarization dependencies were observed in RbV$_{3}$Sb$_{5}$ [Fig.~\ref{fig:4}(c) and~\ref{fig:4}(d)], suggesting that its CDW order parameter closely resembles that of CsV$_{3}$Sb$_{5}$. Remarkably, KV$_{3}$Sb$_{5}$ exhibits significantly different amplitude mode spectra compared to the other two compounds although all three compounds consistently manifest $E_{2g}$ phonon splitting. In the same wavenumber range, KV$_{3}$Sb$_{5}$ displays only a single amplitude mode, which demonstrates a twofold angular intensity pattern [Fig.~\ref{fig:4}(e) and~\ref{fig:4}(f)]. This striking difference strongly suggests that KV$_{3}$Sb$_{5}$ possesses a fundamentally distinct CDW order parameter compared to CsV$_{3}$Sb$_{5}$ and RbV$_{3}$Sb$_{5}$, as the CDW amplitude modes reflect the nature of CDW order parameters. We note that although KV$_{3}$Sb$_{5}$ exhibits an amplitude mode spectrum that differs pronouncedly from those of the Cs and Rb compounds, it nonetheless displays rotational symmetry breaking toward the $C_{2}$ symmetry.

The ubiquitous phonon anomaly accompanied by distinct CDW amplitude modes across \textit{A}V$_{3}$Sb$_{5}$ compounds can be interpreted in two different ways. First, the charge order structure responsible for the rotational symmetry breaking in KV$_{3}$Sb$_{5}$ may differ from that in the Cs and Rb compounds. The material dependence of the amplitude mode spectra captured in our ARPRS measurements bear a strong resemblance to those associated with a 4$a_{0}$ charge stripe order---present in CsV$_{3}$Sb$_{5}$ and RbV$_{3}$Sb$_{5}$ but notably absent only in KV$_{3}$Sb$_{5}$~\cite{zhao_cascade_2021,li_small_2023,xing_optical_2024}. From this perspective, the observed $E_{2g}$ phonon splitting and the consistent CDW amplitude mode spectra in CsV$_{3}$Sb$_{5}$ and RbV$_{3}$Sb$_{5}$ may originate from the 4$a_{0}$ charge stripe order, whereas the phonon splitting and the distinct amplitude mode spectrum in KV$_{3}$Sb$_{5}$ may arise from a different rotational symmetry breaking mechanism. An alternative possible interpretation is that the charge order structure responsible for the $E_{2g}$ phonon splitting is the same across Cs, Rb, and K compounds, but the difference lies solely in their CDW order parameters resulting in the distinct CDW amplitude modes in KV$_{3}$Sb$_{5}$. In this context, the $E_{2g}$ phonon splitting consistently observed in all of \textit{A}V$_{3}$Sb$_{5}$ may originate from a high-temperature rotational symmetry breaking that precedes the onset of the 4$a_{0}$ stripe order, as previously reported in all $\textit{A}$V$_{3}$Sb$_{5}$ compounds via optical birefringence and scanning tunneling microscopy~\cite{xu_three-state_2022,li_rotation_2022,li_unidirectional_2023}. Notably, our magneto-Raman measurements reveal no magnetic field dependence in the P$_{1}$ phonon frequency anomaly (see Supplemental Material). This observation aligns with the previous report that high-temperature rotational symmetry breaking is insensitive to external magnetic fields~\cite{li_rotation_2022}. However, to further refine these interpretations, a deeper understanding is needed of how different forms of CDW order parameters can emerge within the same charge order structure. Therefore, comprehensive investigations into why a distinct symmetry-breaking phenomenon appears only in KV$_{3}$Sb$_{5}$ could provide insights that substantiate these scenarios. For example, the missing amplitude modes observed in KV$_{3}$Sb$_{5}$ could be attributed to disorder effect, and exploring the material-dependent physical properties from the perspective of disorder may offer a possible route to resolving the long-standing puzzle of \textit{A}V$_{3}$Sb$_{5}$ family \cite{wu_charge_2022,wilson_av3sb5_2024}.

Our effective Hamiltonian model, constructed to understand the microscopic origin of the observed $E_{2g}$ phonon splitting and multiple amplitude modes, highlights that $C_{3}$ symmetry breaking can induce not only the lifting of the $E_{2g}$ degeneracy but also the emergence of multiple amplitude modes and their $\theta$-dependent Raman intensities (technical details are provided in the Supplemental Material). Furthermore, by imposing $C_{3}$-symmetry breaking in our ab-initio density-functional perturbation theory (DFPT) calculations using the staggered inverse Star of David (iSoD) structure characterized by orthorhombic $D_{2h}$ symmetry, we obtain a splitting of the $E_{2g}$ phonon mode whose magnitude is comparable to the experimentally observed value (see Supplemental Material). Therefore, our theoretical investigations demonstrate that rotational-symmetry breaking in the CDW phase is the key factor underlying the observed $E_{2g}$ phonon splitting and the emergence of multiple amplitude modes. Indeed, previous Raman and ultrafast pump–probe experiments have shown that the $E_{2g}$ phonon splitting emerges in a staggered iSoD state \cite{jin__2024,deng_revealing_2025}. However, our ARPRS experiments on split $E_{2g}$ phonons assigns the irreducible representations $A_{g}$ and $B_{1g}$ in reversed order compared to the DFPT calculations for the staggered iSoD structure (Supplemental Material). This discrepancy could arise from limitations in the employed pseudopotentials or other computational parameters, or it may indicate that staggered iSoD may not be the correct charge order structure. Therefore, considering other or more complex superstructure beyond this charge order structure and investigating the symmetry indicator is essential not only for explaining our ARPRS results, but also for achieving a proper understanding of the ground state charge order structure in \textit{A}V$_{3}$Sb$_{5}$.

Our ARPRS experiments on \textit{A}V$_{3}$Sb$_{5}$ uncover a unbiquitous phonon anomaly, while the excitation spectra of their CDW order parameters exhibit pronounced material-dependence. These findings establish a foundation for unraveling distinct behaviors in lattice and charge sectors and uncharted material-specific electronic symmetry breaking phenomena. This spectroscopic investigation may sets the stage for elucidating the origin of the absence of phonon softening in these materials~\cite{li_observation_2021,xie_electron-phonon_2022,chen_absence_2025}, a hallmark that distinguishes \textit{A}V$_{3}$Sb$_{5}$ from many other CDW systems~\cite{weber_optical_2013,le_tacon_inelastic_2014,oh_b1g-phonon_2021}. Hence, comprehensive theoretical and experimental studies of collective charge excitations of charge and lattice degrees of freedom will be crucial for revealing the emergent electronic properties in $\textit{A}$V$_{3}$Sb$_{5}$ kagome superconductors. 

\textit{Acknowledgments}---We appreciate discussions with R.~M.~Fernandes, P.~J.~W.~Moll, and L.~Wu. This material is based upon work supported by the National Science Foundation under Grant No.~DMR-2405560. The cryomagnet used in this work (OptiCool, Quantum Design) was acquired with support from the Air Force Office of Scientific Research under the Defense University Research Instrumentation Program (DURIP) Grant FA9550-22-1-0130. We also acknowledge support by the Institute of Applied Physics of Seoul National University. S.D.W.~and A.C.S.~gratefully acknowledge support via the UC Santa Barbara NSF Quantum Foundry funded via the Q-AMASE-i program under award DMR-1906325. Z.W.~acknowledges the support of the U.S. Department of Energy, Basic Energy Sciences Grant No.~DE-FG02-99ER45747. This work is supported by the Deutsche Forschungsgemeinschaft (DFG, German Research Foundation) through Project-ID 258499086--SFB 1170 and through the Würzburg-Dresden Cluster of Excellence on Complexity and Topology in Quantum Matter--ct.qmat Project-ID 390858490--EXC 2147.


%

\end{document}